\documentclass[sigconf]{acmart}
\usepackage{alltt}
\usepackage{graphicx}
\usepackage{balance}  % for  \balance command ON LAST PAGE  (only there!)
\usepackage{hyperref}

\renewcommand\footnotetextcopyrightpermission[1]{} % removes footnote with conference information in first column
\copyrightyear{2019}
\acmYear{2019}
\setcopyright{none}
\acmConference[aiDM '19]{aiDM '19: Second International Workshop on Exploiting Artificial Intelligence Techniques for Data Management}{July 5, 2019}{Amsterdam, NL}
\acmBooktitle{aiDM '19: Second International Workshop on Exploiting Artificial Intelligence Techniques for Data Management, July 5, 2019, Amsterdam, NL}
\acmPrice{15.00}
\acmDOI{}
\acmISBN{}

\begin{document}

% ****************** TITLE ****************************************

\title{Column2Vec: Structural Understanding via Distributed Representations of Database Schemas}

% ****************** AUTHORS **************************************

\author{Michael J. Mior, Alexander G. Ororbia}
\email{mmior@cs.rit.edu, ago@cs.rit.edu}
\affiliation{
   \institution{Rochester Institute of Technology}
   \streetaddress{102 Lomb Memorial Drive}
   \city{Rochester}
   \state{New York}
   \postcode{14623-5608}
}
     
\date{18 March 2019}

\begin{abstract}
We present Column2Vec, a distributed representation of database columns based on column metadata.
Our distributed representation has several applications.
Using known names for groups of columns (i.e., a table name), we train a model to generate an appropriate name for columns in an unnamed table. 
We demonstrate the viability of our approach using schema information collected from open source applications on GitHub.
\end{abstract}

\maketitle

\section{Introduction}

It has become increasingly common for enterprise data management platforms to soak up as much data as possible from a variety of sources.
These ``data lakes'' are often lacking in metadata which makes the structure of stored data challenging to understand.
This limits the usability of this additional data since significant time must be invested by data scientists when adding useful metadata before the data can be analyzed or integrated with other sources.

One common operation is database normalization. When performing normalization, it is common to decompose a single database table into multiple tables based on dependencies which are inferred from the data.
The final schema typically more closely represents logical entities in the underlying data.
Consider the single table below:

\begin{alltt}
  authorID, firstName, lastName, ISBN, title
\end{alltt}

This table contains information on both authors  and books.
A standard normalization algorithm to convert this table into Boyce-Codd Normal Form (BCNF)~\cite{Codd1974} would produce three  tables:

\begin{alltt}
  authorID, firstName, lastName
  ISBN, title
  authorID, ISBN
\end{alltt}

These tables represent information on authors, books, and the relationship between them.
This normal form is useful for data integration tasks but there exists no obvious approach for producing meaningful names for each of these tables.
Currently these names are manually assigned by the database designer.

Our work makes the following contributions:

\begin{enumerate}
\item A semantic embedding of database column names
\item A method for using these embeddings to assign meaningful names to tables containing a given set of column names
\item A metric for evaluating generated table names which shows the usefulness of our prediction. 
\end{enumerate}

\section{Related Work}
\label{sec:related_work}
One of the central motivations behind this work comes from recent progress that has been made in the area of natural language processing (NLP) due to the use of distributed representations.
Distributed representations, or \emph{embeddings}, refer to the mapping of input examples to vectors of values, of possibly lower dimensionality than the input itself.
These embeddings are usually produced through the use of an artificial neural network (ANN).
Each element in one such vector is not necessarily associated with one particular concept, feature, or object, but rather works in tandem with the other elements in the same vector to represent a set of features or concepts that describe the input itself.
The Word2vec family of models, i.e., skip-gram and continuous bag of words, have yielded some of the most widely-used embeddings in NLP, where a simple feed-forward ANN language model is trained on a large collection of documents.
The internal weight vectors, which each map to a particular token, are then used in some subsequent predictive language task, e.g., text classification or chunking.

Other variants have been proposed since Word2vec's initial public release, such as GloVe~\cite{Pennington14}, ELMo~\cite{Peters2018}, and BERT~\cite{Devlin2018}.
Interestingly enough, these representations can be composed into representations of phrases and sentences, by averaging, summing, or concatenating the embeddings for each of the constituent words, yielding a possible distributed representation of the phrase/sentence itself.
Embeddings have found use in domains even outside of NLP, such as in graph/network representation \cite{grover2016node2vec,narayanan2017graph2vec}, including entity resolution~\cite{Ebraheem2017}, concept modeling~\cite{Li2016,Moody2016,Sherkat2017} and data curation~\cite{Thirumuruganathan2018}.
In short, distributed representations have facilitated the construction of a useful, alternative means of comparing, aggregating, and manipulating the fundamental elements of a data type. 

\begin{figure*} % had to put this here to get this on 2nd page of doc
  \centering 
  \includegraphics[scale=0.55]{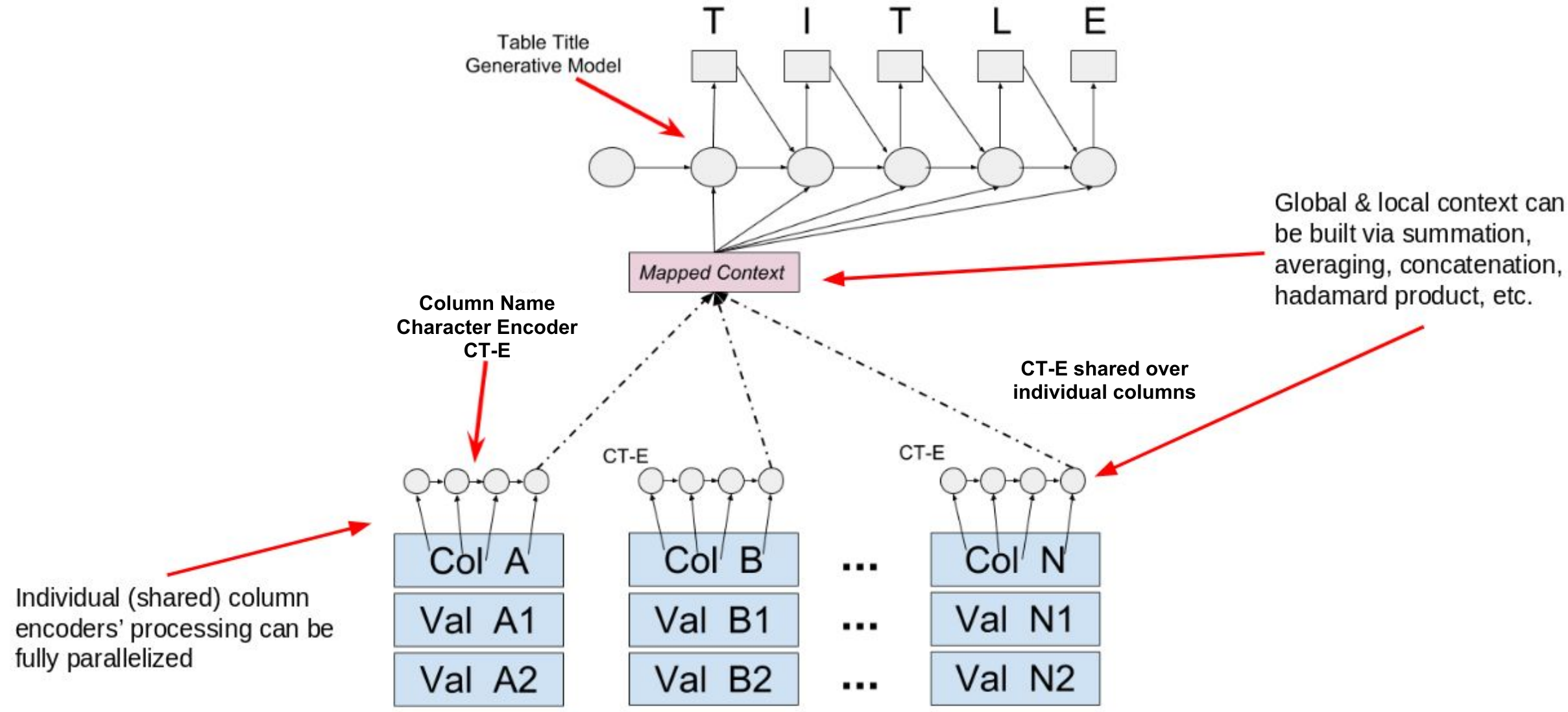}
  \caption{Recurrent generative model of table titles.}\label{fig:name-gen-model}
\end{figure*}

In this work, for tables themselves, we hypothesize that the same potential for discovering aggregated embeddings exists as well -- for each column name, we could find its particular distributed representation, and, after examining the entire set of column names, we could compose a plausible distributed representation of the entire table itself by applying some aggregation operation to the constituent embeddings.
Furthermore, these aggregate table embeddings could be used in the generation of plausible names for the tables themselves.
In this work, we will propose two possible ways in which this might be implemented and empirically explore the effects of one of these.
%This work could be thought of a small step towards automated data pre-processing, with similar work pushing along this direction including...% could connect this to the references/sources below

\section{Models for Table Name Generation}\label{sec:model}
In this section, we will describe two architectures, one simple and one complex, for generating table names.
We will then provide some experimental results for the first model in this study.

\subsection{Model \# 1}
There are two main steps in this approach to generating database table names.
First, we produce a word embedding for each table and column name as discussed in Section~\ref{subsec:vecs}.
Then, as discussed in Section~\ref{subsec:prediction} we use the vectors generated for each column name in a table to predict a meaningful name for the table. 

\subsubsection{Embeddings for Database Columns}\label{subsec:vecs}
To generate word embeddings for Column2Vec, we make use of fastText~\cite{Bojanowski2017}.
This is primarily because fastText allows embeddings to be generated for terms which are  outside of the original vocabulary the model was  trained on.
This is important in our setting since table and  column names contain significant variation and may not appear as an exact match in any existing data.
For example, approximately half of the table names we see in the data set used for our preliminary results in Section~\ref{sec:results} appear only once.

To train the fastText model, we first construct documents based on the names of tables and columns of known database schemas.
For example, one document might consist of the string \texttt{"authors authorID firstName lastName"}.
Training on a collection of such documents allows us to generate embeddings, or \emph{word vectors} for each of the table and column names in our training set.
In addition, fastText also enables us to generate word vectors for terms outside of this vocabulary by looking at subword information.

\subsubsection{Table Name Generation}\label{subsec:prediction}

For our generation task, we aim to take as input a list of column names and assign the most meaningful table name from tables observed in our training set.
Our prediction is based on comparing the similarity of the word vectors for each column vectors.
Since we want to incorporate semantic context from all column names, we need to combine the  individual word vectors generated for each column.
Mikolov et al.~\cite{Mikolov2013} showed that the summation of word vectors is most likely to produce a vector for a term which is semantically similar to the composition of each term.
In our case, we expect the sum of the word vectors associated with a set of columns will produce a vector which is close (in vector space) to a word vector representation of a viable table name.

To produce a table name that might be suitable for the given set of columns, we use the nearest neighbor based on cosine similarity.
That is, after summing the individual column vectors, we select the table name which corresponds to the word vector most similar to this summed vector.
Currently our model considers/examines only the single closest vector.
However, in future work, we intend to investigate the case where $k$ nearest closest neighbors are considered for further evaluation.
%in the future we expect that considering the $k$ closest and evaluating a more complex objective function against these $k$ possible choices may produce better results.

\subsection{Model \# 2}
Another approach we propose is based on a recurrent neural architecture that learns a generative model of table names, conditioned on a distributed representation of table metadata.
A graphical depiction of a basic version of this model can be found in Figure~\ref{fig:name-gen-model}.

The generative model can be decomposed into three general modules: 1) a column name encoder/processor, 2) a table representation aggregation function, and 3) a conditional name generator.
All three functions could be as simple as feed-forward network models or as complex and powerful as a gated recurrent neural network (RNN), such as Long Short Term Memory~\cite{hochreiter1997long} or $\Delta$-RNN~\cite{ororbia2017learning}.
We envision the column name encoder and table name generator to operate at the character symbolic level (or the subword level) to circumvent the problem of an incredibly high-dimensional input/output space that word-level models have to face.

The column name encoder can be a recurrent network that builds a stateful representation of a single column by iteratively processing the underlying characters that compose the word or phrase used to label the column name itself.
This column name encoder would be shared across column names to drastically reduce the number of parameters required in our model and allow it to easily handle a variable number of columns.
A more difficult, but perhaps quite fruitful, extension of this encoder would be to have it process the data values associated with a particular column name as well.
This encoder would be applied to each column name and output a set of $K$ column name representations.

The $K$ embeddings produced by the column name encoder would then be run through an aggregation function, which could itself be a nonlinear multilayer perceptron (MLP) or a simple function such as averaging or (weighted) summation.
This function's primarily role is to create a single, fixed-length representation of a set of column names, or rather, the table of interest itself.
This table representation would then be used to guide a generative model of text, which could be a simple RNN as depicted in Figure~\ref{fig:name-gen-model}.
By focusing on a character or subword-level RNN generative model, we can naturally handle names of variable length and even potentially generate non-standard symbols (such as alphanumeric strings).
The types of names that would be generated by this model would largely depend on the table dataset used to construct the overall model, however, if large enough, the model might be able to produce some interesting, creative table names or even a set of candidates that the human user would finally select from and/or error-correct (providing additional feedback to the model).

Parameters of the entire end-to-end model could be learned using reverse-mode differentiation to optimize an objective such as the negative log likelihood of the name text in the training set.
Since the entire system is soft, or rather, makes use of differentiable nonlinear transformations, calculating gradients would not be difficult, though the path of credit assignment could potentially be long depending on how long some table names and column names are (since in RNN structures, the model parameters are shared over time-steps, so in the case of characters, we apply the RNN parameters once per character).

\section{Preliminary results}\label{sec:results}

To examine the effectiveness of our first model, we used a set of table schemas (column and table names) collected from a crawl of open source repositories on GitHub~\cite{Hoffa2016}.
Files in all repositories were checked for the syntactically valid \texttt{CREATE TABLE} statements and the table and column names were extracted from these statements. 
This resulted in a total of $436,545$ tables combined with the names of each column in the table.
We leave implementation and evaluation of our second model as future work.

\subsection{Data cleaning}

The dataset pulled from GitHub contains a significant amount of test and dummy data which is not useful for our problem.
For example, a table with the name \texttt{bb} and columns \texttt{col1} and \texttt{col2} would be discarded.
We implemented a simple set of rules to filter the extracted information:
(1) trigrams appearing in table or column names appearing only once, eliminating random entries,
(2) names with special characters,
(3) names with a large number of digits, and
(4) names which consist of only two repeated characters (e.g., \texttt{bb}).

\subsection{Model training}
\label{sec:training}
The data was split into 90\% training data with 10\% reserved for testing.
Word vectors were trained using the fastText skip-gram model with each document consisting of the table name and associated column names.
We performed hyper-parameter optimization using the tree-structured Parzen estimator approach described by Bergstra et al.~\cite{Bergstra2011} on a subset of the data.
A k-nearest neighbors model was then trained using the table name word vectors (in the training set) to generate table names as described in Section~\ref{sec:model}.

\subsection{Table name quality}
\label{sec:name_quality}
Table names may consist of multiple words concatenated together.
A table storing blog posts may be called \texttt{blogposts}, but we might consider a suggestion of the name \texttt{posts} to be a reasonable substitute.
We want to be able to identify how well the components of our predicted names match the original.
To identify the components of a name, we first attempt to split names into words via a dynamic programming approach that aims to infer the position of spaces by attempting to maximize the probability of individual word frequencies~\cite{wordninja}.
This is based off of a sample of words from the English Wikipedia corpus.

To evaluate the quality of the generated table names, we define a metric based on the $F_1$ score~\cite{Chinchor92} which combines precision and recall.
We use the words split from the table names as  mentioned above when calculating the $F_1$ score.
However, we also want words which are semantically similar to rank high.
For example, the name \texttt{books} may a suitable alternative to the name \texttt{library}.
To capture these relationships, we use the path similarity from WordNet::Similarity~\cite{Pedersen2004}.
When computing precision and recall, instead of using intersection between the predicted and original names, we calculate \emph{fuzzy} precision and recall using this WordNet similarity metric.

\begin{figure}
  \centering 
  \includegraphics[scale=0.4]{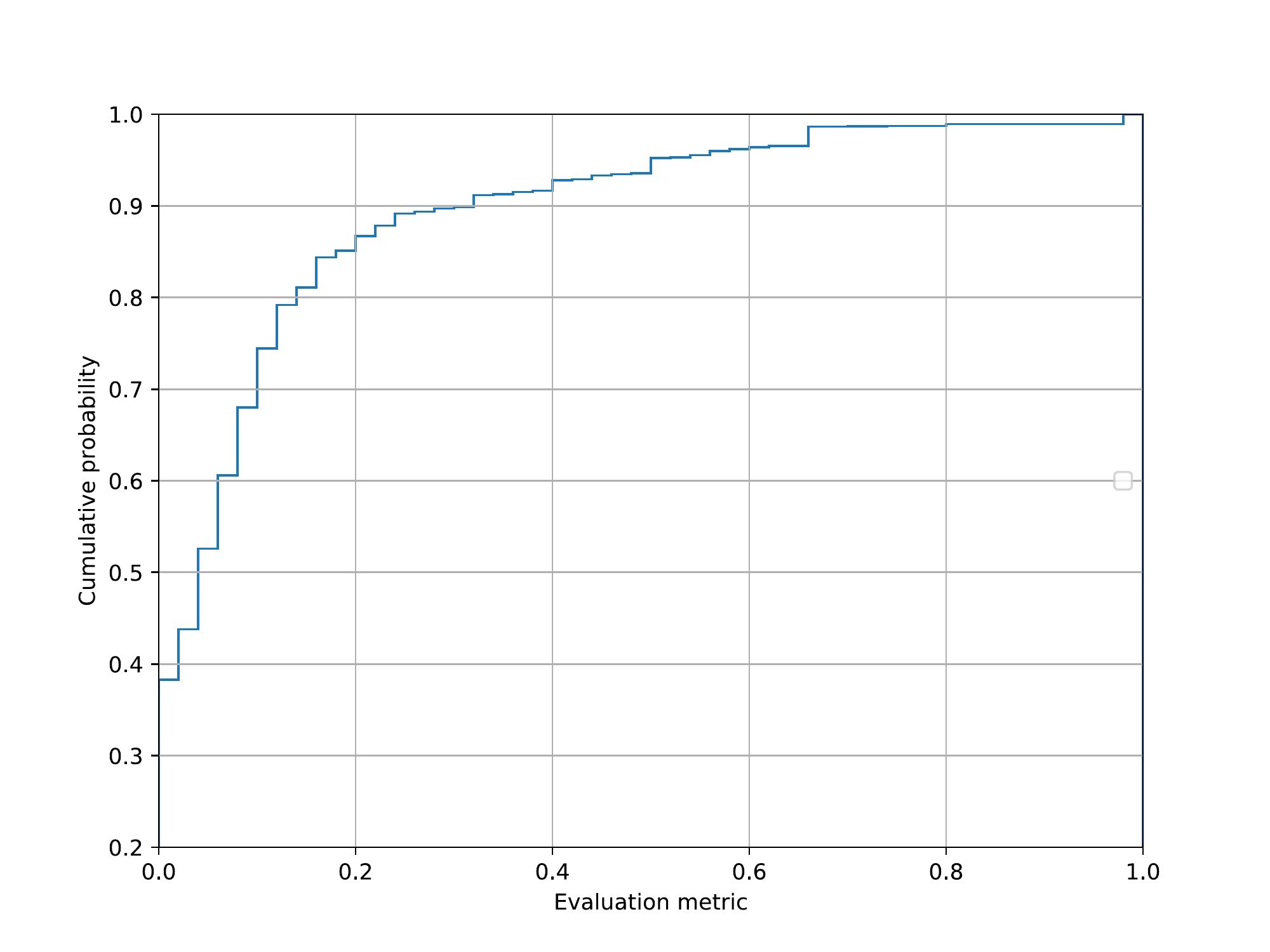}
  \caption{Cumulative distribution of evaluation metric}\label{fig:eval-metric-cdf}
\end{figure}

Figure~\ref{fig:eval-metric-cdf} shows the cumulative distribution of the evaluation metric.
Approximately one third of the predicted table names evaluate to zero.
Of these, roughly 60\% have the same two predicted tables, \texttt{vtp} and \texttt{defaultrequiredlengthncharcolumns}.
Neither of these are semantically meaningful.
Further preprocessing to remove names with limited semantic value will likely improve this result.
We also see that approximately 1\% of the values have the highest possible score of 1.0.
In this case, names differed just in pluralization and punctuation, such as \texttt{recipe\_ingredients} and \texttt{recipeingredient}.
We provide examples of results which fall between these two extremes in the following section.

\balance

\subsection{Table Name Samples}

Table~\ref{tbl:examples} gives examples of the original names assigned to tables, i.e., the ground truth, as well as the predicted name and the value for our evaluation metric.

\begin{table}
  \begin{tabular}{ccc}
    Metric value & Original name & Predicted name \\\hline
    0.11 & tcardfieldoption & ttextboxsettings \\
    0.13 & subscription & oc\_product\_description \\
    0.20 & portfolio & projects \\
    0.25 & campaignchain\_group & userseries \\
    0.33 & comments & content \\
    0.39 & mg\_tag\_properties & forum\_articles\_tags \\
    0.43 & users\_clients & osiam\_client \\
    0.50 & item & symbol \\
    0.56 & useridentity & usercredentials \\
    0.67 & initializedmodules & module \\
    0.71 & artist\_tag\_raw & release\_tag\_raw \\
    0.75 & sales\_order\_item & sales\_order\_line \\
    0.80 & arena\_team & arena\_team\_member \\
    0.86 & inserttesttablefilter & inserttesttable \\
    1.00 & position & positions
  \end{tabular}
  \caption{Example table name predictions}\label{tbl:examples}
\end{table}

Although the examples above suggest that our metric is useful, in the future we intend to perform a more thorough evaluation to determine whether this metric correlates with human judgment.

To better demonstrate our generation/evaluation process, consider predicting a name for a table with columns \texttt{id}, \texttt{calendarid}, \texttt{name}, \texttt{eventdate}, and \texttt{locutionid}.
We generate word vectors for each of these columns and sum them.
When searching for the table name with the closest word vector, we find \texttt{eventdates}.

The original name given to this table was \texttt{holidaydates}.
To evaluate this according to our metric we first split each name into words giving [\texttt{holiday}, \texttt{dates}] and  \texttt{event}, \texttt{dates}].
Based on the WordNet path similarity, \texttt{holiday} and \texttt{event} have a similarity of 0.14.
Since \texttt{dates} matches exactly, we end up with precision and recall values of $P = R = (1.14 / 2) = 0.57$.
Our final metric value is then $2PR / (P + R) = 0.57$.
Note that in this case, we have no indication based on column names that the table specifically refers to holidays.
In the future, we will explore other sources of contextual information which may help produce more accurate results.

\section{Conclusions and Future Work}

Generating meaningful names for tables given only constituent column names is a challenging problem.
Our approach, based on distributed representations, is able to generate meaningful table names given the names of the columns contained in the table.
While our metric does prove useful, additional work is needed to compare it properly against human judgment.

We believe that incorporating information on the data stored in these columns (e.g. data type and value distribution) will make this representation even more useful. %This, however, complicates training data collection.
%This also significantly complicates the collection of useful training data.
In addition to generating tables names, such a representation would likely benefit other data integration tasks, e.g., deciding which tables in a large set may be meaningfully joined together.

\bibliographystyle{ACM-Reference-Format}
\bibliography{column2vec}  % vldb_sample.bib is the name of the Bibliography in this case

\end{document}